\documentclass[floatfix, preprint, onecolumn, 	prd,showpacs, showkeys, preprintnumbers, nofootinbib]{revtex4-1}
\usepackage{graphicx}
\usepackage{epsfig,float}
\usepackage[sort&compress]{natbib}
\usepackage{subfigure}
\usepackage{amsmath}
\usepackage{amsfonts}
\usepackage{cancel}
\usepackage{amssymb}
\usepackage{hyperref}
\usepackage{color}

\makeatletter
\let\oldabs\abs
\def\abs{\@ifstar{\oldabs}{\oldabs*}}
\let\oldnorm\norm
\def\norm{\@ifstar{\oldnorm}{\oldnorm*}}
\makeatother

\begin{document}

\title{The $\Lambda_{b}\rightarrow J/\psi K^{0}\Lambda$ reaction and a hidden-charm pentaquark state with strangeness}

\author{Jun-Xu Lu$^{1}$}
\author{En Wang$^{2}$}
\email{wangen@zzu.edu.cn}
\author{Ju-Jun Xie$^{3,4,5}$}
\email{xiejujun@impcas.ac.cn}
\author{Li-Sheng Geng$^{1,5}$}
\email{lisheng.geng@buaa.edu.cn}
\author{Eulogio Oset$^{3,6}$}
\email{Eulogio.Oset@ific.uv.es}

\affiliation{
$^1$School of Physics and Nuclear Energy Engineering and International Research Center for Nuclei and Particles in the Cosmos, Beihang University, Beijing 100191, China \\
$^2$Department of Physics, Zhengzhou University, Zhengzhou, Henan 450001, China \\
$^3$Institute of Modern Physics, Chinese Academy of Sciences, Lanzhou 730000, China \\
$^4$Research Center for Hadron and CSR Physics, Institute of Modern Physics of CAS and Lanzhou University, Lanzhou 730000, China\\
$^5$State Key Laboratory of Theoretical Physics, Institute of Theoretical Physics, Chinese Academy of Sciences, Beijing 100190, China \\
$^6$Departamento de F\'{\i}sica Te\'orica and IFIC, Centro Mixto Universidad de Valencia-CSIC Institutos de Investigaci\'on de Paterna, Aptdo. 22085, 46071 Valencia, Spain}

\begin{abstract}
We study the $\Lambda_{b}\rightarrow J/\psi K^{0}\Lambda$ reaction considering both the $K^{0}\Lambda$ interaction with its coupled channels and the $J/\psi\Lambda$ interaction. The latter  is described by taking into account the fact that there are predictions for a hidden-charm state with strangeness that couples to $J/\psi\Lambda$. By using the coupling of the resonance to $J/\psi\Lambda$ from these predictions we show that a neat peak can be observed in the $J/\psi\Lambda$ invariant mass distribution, rather stable under changes of unknown magnitudes. In some cases, one finds a dip structure associated to that state, but a signal of the state shows up in the $J/\psi$ spectrum. \end{abstract}


\date{\today}

\maketitle
\section{Introduction}
Interests on pentaquark states have been reignited with the latest LHCb observation of the $P_c(4380)$ and $P_c(4450)$ in the $\Lambda_b \to J/\psi p K^-$ decay~\cite{Aaij:2015tga}.  Since both states
are observed in the $J/\psi p$ invariant mass distributions, their minimum quark components should be $c\bar{c}uud$. Due to the large charm quark mass, these states seem to be 
no less exotic than the $\theta^+(1540)$, 
whose first claim
was made by the LEPS collaboration~\cite{Nakano:2003qx} (see also update in Ref.~\cite{new}) but 
later on found to be rather controversial experimentally ~\cite{Hicks:2012zz}.  A 
reanalysis of Ref.~\cite{new} was done in Refs.~\cite{alber1,alber2},
showing that the peak observed was a consequence of an artificial method 
used in Ref.~\cite{new} to determine invariant masses with an incomplete kinematics. The apparently exotic nature
of the hidden-charm pentaquark states has aroused a lot of theoretical interests to interpret their nature.  For instance, they have
been proposed to be meson-baryon molecules~\cite{Chen:2015loa,Roca:2015dva,He:2015cea,Huang:2015uda,Meissner:2015mza,Xiao:2015fia,Chen:2015moa,Eides:2015dtr,Yang:2015bmv,Huang:2015uda}, diquark-diquark-antiquark pentaquarks~\cite{Maiani:2015vwa,Anisovich:2015cia,Ghosh:2015ksa,Wang:2015epa,Wang:2015ixb}, compact diquark-triquark pentaquarks~\cite{Lebed:2015tna,Zhu:2015bba},  $\bar{D}$-soliton states~\cite{Scoccola:2015nia}, genuine multiquark states~\cite{Mironov:2015ica,Gerasyuta:2015djk},
and kinematical effects related to the so-called triangle singularity~\cite{Guo:2015umn,Liu:2015fea,Mikhasenko:2015vca}.~\footnote{See Refs.~\cite{Wu:2015nhv,Stone:2015iba}, for a nice summary of theoretical and experimental activities, a recent extended review can be seen in Ref.~\cite{Chen:2016qju}}

Clearly, not all of the theoretical interpretations are consistent with each other. The molecular interpretations cannot easily accommodate  simultaneously the two states with opposite
parity, which is not the case for QCD sum rules. The fact that the Fock components in the wave function of a hadronic state are themselves not observable complicated further the discussion. 
As such, it has been stressed that to differentiate the various structures it is important to study the production and decay patterns of these pentaquark states,  e.g.,  the weak decays of bottom baryons~\cite{Li:2015gta,Cheng:2015cca}, photo-productions~\cite{Wang:2015jsa,Kubarovsky:2015aaa,Karliner:2015voa}, the $\pi^- p \to J/\psi n$ reaction~\cite{Lu:2015fva},  elastic and inelastic $J/\psi N$ cross sections~\cite{Xiao:2015fia}, and the strong decays of these states~\cite{Wang:2015qlf}. Searches for the counterparts of these two states, such as its strangeness counterpart~\cite{Chen:2015sxa,feijoonew}, will offer new insight into their true nature as well.

In such a context, it is important to note that as proposed in a recent work~\cite{Wang:2015pcn}, a reanalysis of the $\Lambda_b\rightarrow J/\psi \pi^- p$ decay is very important. In the $\pi^- p$ invariant mass 
distribution of this decay process, an enhancement followed by a dip can be seen~\cite{Aaij:2014zoa}. A careful study of this decay showed that the structure is consistent with the existence of the $P_c(4450)$, which
couples to $J/\psi p$ and coupled channels in $s$-wave~\cite{Wang:2015pcn} .  It should be mentioned
that the possibility that  the peak seen in Ref.~\cite{Aaij:2014zoa} could correspond to the $P_c(4450)$ was already noted in Ref.~\cite{Burns:2015dwa}, but the possibility that it would correspond to a resonance was not mentioned in the
original experimental paper~\cite{Aaij:2014zoa} probably because of the relatively low
statistics.

In Ref.~\cite{Wang:2015pcn}, the experimental structure close to 4450 MeV was reasonably described. The peak  and dip structure was explained as the interference between $s$-wave $\pi^- p$ (and coupled channels) and 
$J/\psi p$ interactions. If this is confirmed by the ongoing LHCb analysis, it will give further support to the nature of the $P_c(4450)$ proposed in Refs.~\cite{Wu:2010vk,Wu:2010jy}. It should be mentioned that the $\Lambda_b\rightarrow
J/\psi \pi^- p$ decay was also discussed in Refs.~\cite{Burns:2015dwa,Hsiao:2015nna,Cheng:2015cca}.  

Even before their observations, the existence of hidden-charm pentaquark states has been speculated in Refs.~\cite{Wu:2010jy,Wu:2010vk,Yang:2011wz,Xiao:2013yca,Uchino:2015uha,Karliner:2015ina,Garzon:2015zva,Wang:2011rga,Yuan:2012wz,Huang:2013mua}.  One should note that most theoretical approaches
predicted the existence of the $P_c$ counterparts. For instance, in the unitary approach of Ref.~\cite{Wu:2010vk}, in addition to an isospin 1/2 and strangeness zero state, 
two more states are predicted in the isospin zero and strangeness $-1$ sector.~\footnote{
It is interesting to mention the  recent study on the existence of a $s\bar{s}uud$ state~\cite{Lebed:2015dca}.}  The impact of the existence of such (a) state(s) is recently discussed in Refs.~\cite{Chen:2015sxa,feijoonew}, showing that it is possible to
observe it by a careful study of the $\Xi^-_b\rightarrow J/\psi K^-\Lambda$ decay and the $\Lambda_b \to J/\psi \eta  \Lambda$ decay, regardless of its true nature, as long as it exists and couples in $s$-wave to $J/\psi\Lambda$ with a reasonable strength. 
In this present work, we extend such an idea to explore the $\Lambda_b\rightarrow J/\psi K^0\Lambda$ decay. This decay channel is particularly interesting in view of the
structure, which could well be the $P_c(4450)$, observed in the $\Lambda_b\rightarrow J/\psi \pi^- p$ invariant mass distribution~\cite{Wang:2015pcn,Aaij:2014zoa}, because $\pi^-p$ and $K^0\Lambda$ are coupled channels themselves.

This paper is organized as follows. In Sec.~II, we briefly describe  the mechanism of the weak $\Lambda_b\rightarrow J/\psi K^0\Lambda$ decay. 
Numerical results and discussions, focusing on the effects of variation of the mass, width, and coupling strength to $J/\psi\Lambda$ of the 
hidden-charm pentaquark state as well as the contributions of other $N^*$ states in addition to the $N^*(1535)$,  are presented in Sec.~III, followed by a short
summary in Sec.~ IV.

\section{Formalism}
\label{sec:formalism}
\begin{figure}
  \centering
  \includegraphics[width=0.8\textwidth]{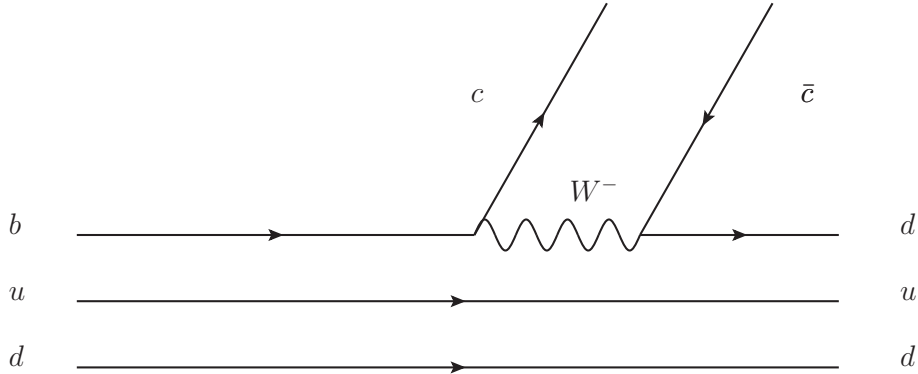}\\
  \caption{Diagrammatic representation of the $\Lambda_{b}\rightarrow J/\psi + dud$ process.}\label{Fig1}
\end{figure}

The $\Lambda_{b}\rightarrow J/\psi K^{0}\Lambda$ decay follows at quark level a process similar to the one described in the $\Lambda_{b}\rightarrow J/\psi K^{-}p$ decay~\cite{Roca:2015dva,Roca:2015tea}, the $\Xi_{b}\rightarrow J/\psi K\Lambda$ decay~\cite{Chen:2015sxa},  the $\Lambda_b \to J/\psi \eta \Lambda$ decay~\cite{feijoonew},
the $\Lambda_b\rightarrow J/\psi K\Xi $ decay~\cite{Feijoo:2015cca}, and the $\Lambda_{b}\rightarrow J/\psi \pi^{-}p$ decay~\cite{Wang:2015pcn}. In particular, the $\Lambda_{b}\rightarrow J/\psi K\Lambda$ decay appears as a coupled channel of the $\Lambda_{b}\rightarrow J/\psi \pi^{-}p$. Diagrammatically this decay is depicted in Fig.~\ref{Fig1},  where the $c\bar{c}$ pair combines to give the $J/\psi$ and the $dud$ quarks will recombine to provide the $K^0\Lambda$ state at the end.

It is interesting to note that the mechanism of Fig.~\ref{Fig1} has the original $u, d$ quarks acting as spectators. This is relevant, because, since the $\Lambda_{b}$ has $I=0$, then the $ud$ original quarks have $I=0$, and so will  they have in the final state if they are spectators. This means that the final state will have the isospin of the ``upper" $d$-quark, and hence $I=1/2$. This picture is strongly supported by the experiment that does not show the smallest trace of a $\Delta(1232)$ in the $\pi^{-}p$ mass distribution~\cite{Aaij:2014zoa}. In the related $\Lambda_{b}\rightarrow J/\psi K^{-}p$ decay, the $ud$ original quarks of the $\Lambda_{b}$ also act as spectators and in this case the extra $s$-quark leads to final $I=0$, $\Lambda^{*}$ baryon states, which were the only ones showing up in the $K^{-}p$ mass distribution in the experimental analysis of Ref.~\cite{Aaij:2015tga}.

In addition, we assume that the $ud$ quark pair of the original $\Lambda_{b}$ goes  into the final baryon upon hadronization of the three quarks produced in the first stage. Other possibilities involve transfer of one of these quarks to the final mesons and a large momentum transfer that strongly  suppresses such mechanisms~\cite{Roca:2015tea,Miyahara:2015cja}.

\begin{figure}
  \centering
  \includegraphics[width=0.8\textwidth]{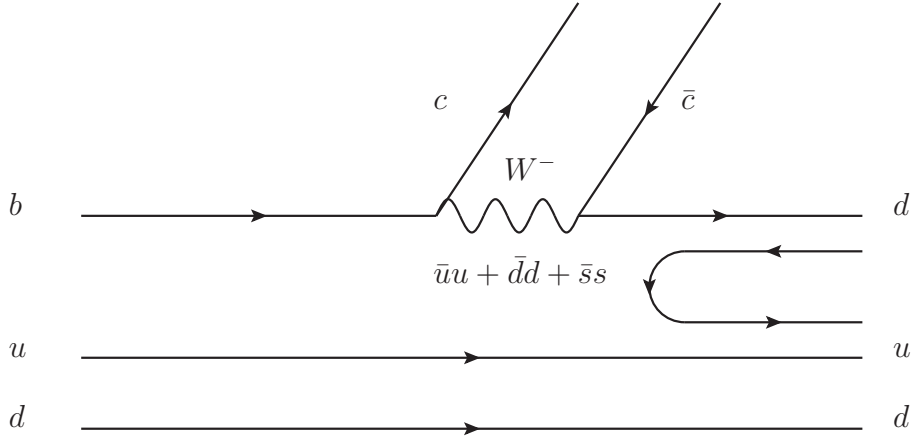}\\
  \caption{Mechanism of Fig.\ref{Fig1} with the hadronization by creation of the $\bar{u}u+\bar{d}d+\bar{s}s$ pair.}\label{Fig2}
\end{figure}

In order to get a meson-baryon pair from the final three quarks of $uud$ we must proceed with hadronization creating a $\bar{q}q$ pair with the quantum numbers of the vacuum. This process must involve the ``upper" $d$-quark in Fig.~\ref{Fig1} because the $K^0\Lambda$ system will be in s-wave in the final state and will have negative parity. Since the quarks of the $I = 0$ $ud$
pair of the $\Lambda_b$ are in the ground level, they will have positive parity and, hence,  the ``upper" $d$-quark must be the one that carries negative parity being in an excited $L=1$ state. Since in the kaon system this quark will again be in the ground state, the hadronization must involve this quark. As a result, the hadronization proceeds as shown in Fig.~\ref{Fig2}

The exercise of finding which meson-baryon components come out from this hadronization was already done in Ref.~\cite{Wang:2015pcn}, and  the combination found  is
\begin{equation}\label{hadronization}
  |H\rangle = \pi^{-}p - \frac{1}{\sqrt{2}}\pi^{0}n + \frac{1}{\sqrt{3}}\eta n + \sqrt{\frac{2}{3}} K^{0} \Lambda.
\end{equation}
It is worth noting that the $\pi N$ component  has indeed $I=1/2$, as well as the $\eta n$ and $K^0\Lambda$ components.

\begin{figure}[t]
\includegraphics[width=0.48\textwidth]{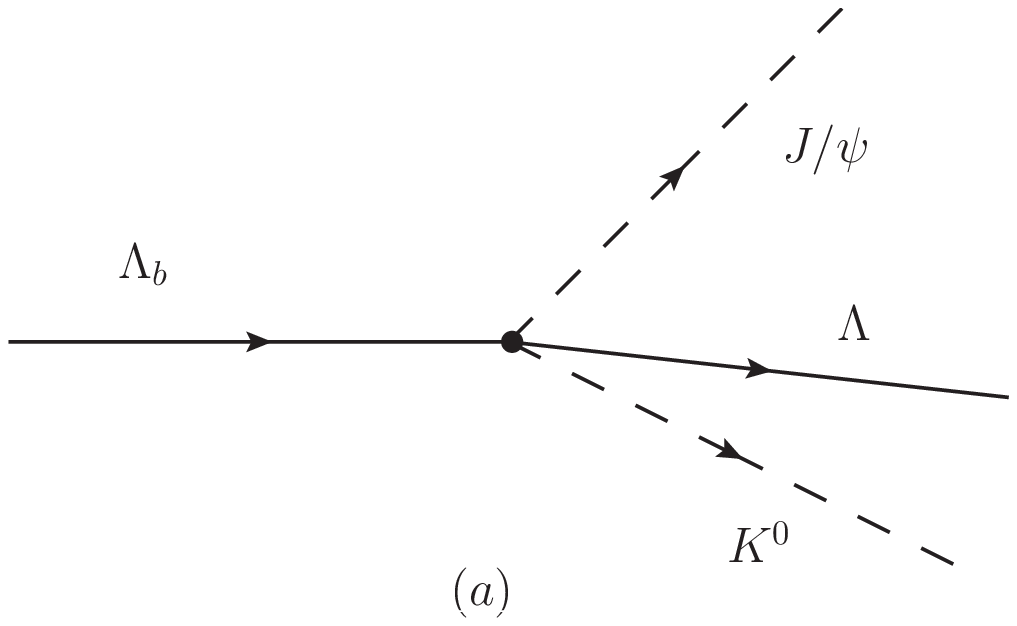}
\includegraphics[width=0.48\textwidth]{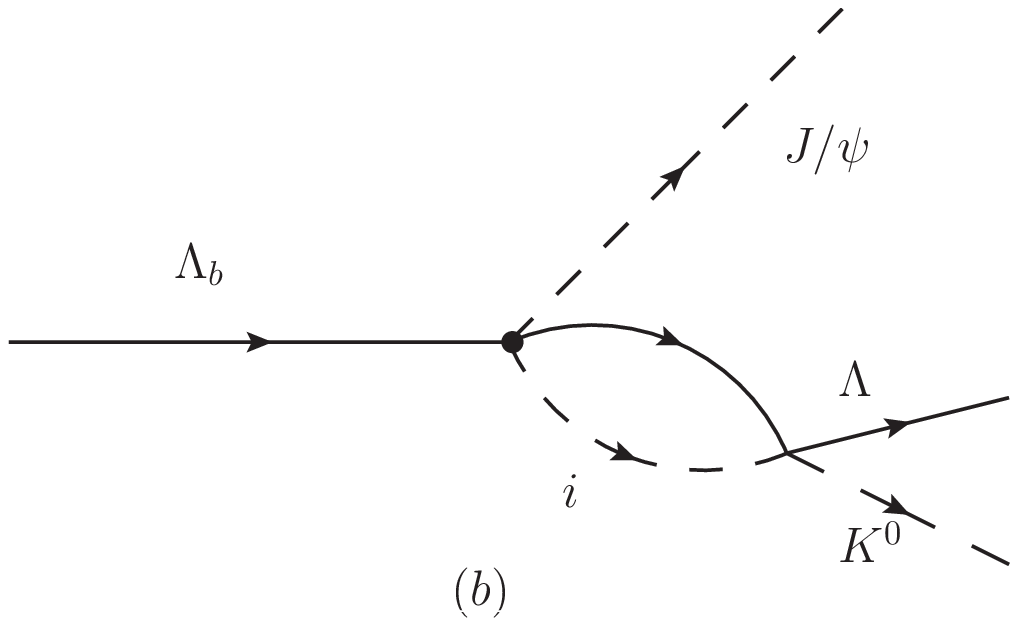}
\caption{Final state interaction of the meson baryon components: (a) tree level contribution, (b) rescattering. }
\label{Fig3}
\end{figure}

The next step is to consider the final state interaction of these channels to give $K^{0}\Lambda$ at the end. This means that, to have $K^{0}\Lambda$ in the final state, we can have it by direct production in the $|H\rangle$ production, by rescattering of $K^{0}\Lambda\rightarrow K^{0}\Lambda$, or by primary production of $\pi N$ or $\eta N$ that go to $K^{0}\Lambda$ after rescattering. This is depicted in Fig.~\ref{Fig3} and analytically this is taken into account by means of
\begin{equation}\label{amplitude1}
  T=V_{p}[h_{K^{0} \Lambda}+\sum_{i}h_{i}G_{i}(M_{K^{0} \Lambda})t_{i\rightarrow K^{0} \Lambda}(M_{K^0\Lambda})],
\end{equation}
where $i\equiv \pi^{-}p, \pi^{0}n, \eta n, K^{0}\Lambda$, and
\begin{equation}\label{h}
  h_{\pi^{-}p}=1, h_{\pi^{0}n}=-\frac{1}{\sqrt{2}},h_{\eta n}=\frac{1}{\sqrt{3}}, h_{K^{0}\Lambda}=\sqrt{\frac{2}{3}}.
\end{equation}

The $G_{i}$ function in Eq.~(\ref{amplitude1}) is the loop function of a meson-baryon and $t_{i\rightarrow K^{0} \Lambda}$ are the scattering matrices of the coupled channels. We take the matrix $t_{i\rightarrow K^{0} \Lambda}$ in $s$-wave from the chiral unitary approach of Ref.~\cite{Inoue:2001ip}, and the loop function $G_i$ from Ref.~\cite{Wang:2015pcn}. Some relevant discussions will be given in the end of this section. 
The factor $V_p$ in Eq. ~(\ref{amplitude1}), which we take as constant (no invariant mass dependence), accounts for the weak and hadronization form factors, which are rather smooth in the limited region of invariant masses that we study \cite{Kang:2013jaa,Daub:2015xja}.
 
We take the philosophy that the process proceeds involving the smallest possible orbital angular momentum in the vertices. In this case, the structure of the $V_p$ vertex is $\vec \sigma \cdot \vec{\epsilon}$, with $\vec{\epsilon}$ the polarization of $J/\psi$ (see appendix). We have $L'=0$, and both the $K^0\Lambda$ and $J/\psi\Lambda$ systems are in total angular momentum $J=1/2$. This must be the case to match the spin of the $\Lambda_b$ (parity is no problem in weak interaction) since the $K^0$ carries zero spin. The spin of $J/\psi\Lambda$ could in principle be $1/2$, $3/2$, but only $1/2$ is allowed and in fact the $\vec \sigma \cdot \vec{\epsilon}$ operator projects the $J/\psi\Lambda$ system in $J=1/2$ as shown in Ref.~\cite{Garzon:2012np} (see Appendix B of this reference).
If we want to have $J=3/2$ for $J/\psi\Lambda$, we need a $p$-wave in the vertex and this will be studied in Sec.~\ref{sec:32}. 

By means of the formalism so far described we could get the $K^{0}\Lambda$ invariant mass distribution in this reaction. Yet, our purpose is to see if in this reaction one could see a signal of a hidden-charm strange molecular state of $J^P=1/2^-,3/2^{-}$ predicted in Refs.~\cite{Wu:2010jy,Wu:2010vk}, mostly made of $\bar{D}^{*}\Xi_{c}^{'}$. The state, degenerate in spin, was produced in $s$-wave, and one of the coupled channels was $J/\psi\Lambda$. 
The  relevant information needed here is the coupling of the resonance to the $J/\psi \Lambda$ system, $g_{J/\psi \Lambda}$, by means of which we can construct the $J/\psi \Lambda \rightarrow J/\psi \Lambda$ amplitude in $s$-wave as
\begin{equation}\label{amplitude2}
  t_{J/\psi \Lambda\rightarrow J/\psi \Lambda}=\frac{g_{J/\psi\Lambda}^2}{M_{J/\psi \Lambda}-M+{\rm i}\Gamma/2}.
\end{equation}

In Refs.~\cite{Wu:2010jy,Wu:2010vk} the mass obtained was $4547-6.4i$, but in Ref.~\cite{Chen:2015sxa} we made a guess of the mass which would be the one of the $P_{c}(4450)$ seen in the $\Lambda_{b}\rightarrow J/\psi K^{-}p$ decay plus an average extra mass of $\Delta M\backsimeq$ 200 MeV due to the mass difference between the $s$-quark and the $u,d$ ones. Thus, tentatively, we begin with a mass $M$ in Eq.~(\ref{amplitude2}) $M\backsimeq 4650$ MeV, but we will change this mass to see what happens for other values. Similarly, the width obtained in Refs.~\cite{Wu:2010jy,Wu:2010vk} was $\Gamma \backsimeq13$ MeV, 
we use $\Gamma=10$ MeV but we will also see what happens for different values of the width.

\begin{figure}
  \centering
  \includegraphics[width=0.5\textwidth]{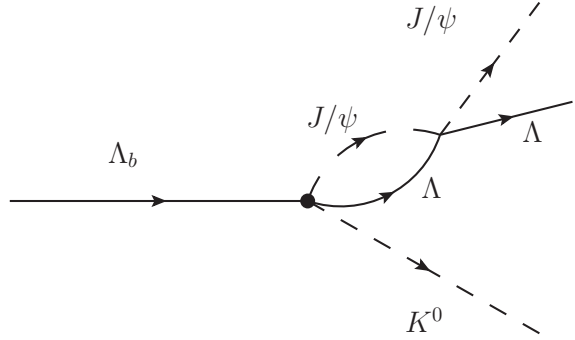}\\
  \caption{Final state interaction of the $J/\psi\Lambda$ state.}\label{Fig4}
\end{figure}

In order to study the effect of this hidden-charm strange state in the process we take into account the $J/\psi \Lambda$ interaction as depicted in Fig.~\ref{Fig4}. Therefore, the final amplitude $\mathcal{M}$  for the process is given by
\begin{eqnarray}\label{amplitude3}
    \mathcal{M}(M_{J/\psi\Lambda},M_{K^0\Lambda}) &  =& V_{p} \Big[h_{K^{0} \Lambda}+\sum_{i}h_{i}G_{i}(M_{K^{0} \Lambda})t_{i\rightarrow K^{0} \Lambda}(M_{K^0\Lambda})\nonumber \\
    &&+h_{K^{0} \Lambda}G_{J/\psi \Lambda}(M_{J/\psi \Lambda})t_{J/\psi \Lambda\rightarrow J/\psi \Lambda}(M_{J/\psi\Lambda})\Big]\nonumber\\
       &=&T_\mathrm{tree}+T_{K^0\Lambda}+T_{J/\psi\Lambda}
\end{eqnarray}
where we have explicitly spelled out the arguments $(M_{K^{0} \Lambda},M_{J/\psi \Lambda})$ of the different $G$ and $t_{ij}$ functions. The $t_{J/\psi \Lambda\rightarrow J/\psi \Lambda}$ amplitude is given by Eq.~(\ref{amplitude2}) and the value of $g_{J/\psi \Lambda}\backsimeq 0.5$ from Refs.~\cite{Wu:2010jy,Wu:2010vk}. We will also investigate what happens under changes of this value. The $G_{J/\psi\Lambda}$ function is  given in Refs.~\cite{Wu:2010jy,Wu:2010vk} using dimensional regularization and the parameters used are $\mu=1000$ MeV and $a_\mu=-2.3$.

As mentioned above, both the $K^0\Lambda\to i$ transitions and $J/\psi\Lambda \to J/\psi\Lambda$ are in $s$-wave, and we can sum them coherently in Eq.~(\ref{amplitude3}) with no spin nor angular structure. 
But this implies that both $J/\psi\Lambda$ and $K^0\Lambda$ are in $J^P=1/2^-$, as we discussed above.

Since the amplitude $\mathcal{M}$ of Eq.~(\ref{amplitude3}) depends on the two invariant masses we will use the two dimensional mass distribution of \cite{Agashe:2014kda}
\begin{equation}\label{distribution}
    \frac{{\rm d}^{2}\Gamma}{{\rm d}M_{J/\psi \Lambda}^{2}{\rm d}M_{K^0 \Lambda}^{2}}=\frac{1}{(2\pi)^{3}}\frac{4M_{\Lambda_{b}}M_{\Lambda}}{32M_{\Lambda_{b}}^{3}}\bar{\sum}\sum|\mathcal{M}(M_{J/\psi \Lambda}, M_{K^0 \Lambda})|^{2},
\end{equation}
where the factor $4M_{\Lambda_{b}^{0}}M_{\Lambda}$ is due to our normalization of the spinors, i.e., $\bar{u}u=1$. By integrating Eq.~(\ref{distribution}) over one or the other invariant mass we obtain the mass distribution of $J/\psi \Lambda$ or $K^{0}\Lambda$.

One technical point is that following Ref.~\cite{Wang:2015pcn} we take the $G_{i}$ loop functions in Eq.~(\ref{amplitude3}) their values using a cut off of $|\vec{q}_\mathrm{max}|=1200$ MeV, instead of those used in Ref.~\cite{Inoue:2001ip}, where dimensional regularization was used that required some odd subtraction constants to account for missing channels that were later identified as $\rho N$ and $\pi \Delta$~\cite{Garzon:2012np}. Since the channels in the loop of Fig.~\ref{Fig3} only contain the $\pi N$, $\eta N$, and $K\Lambda$ channels, the use of the $G_i$ function of Ref.~\cite{Garzon:2012np} was not justified and we use instead the more natural cut off regularization. We must note, however, that this choice does not affect the behavior of the $J/\psi \Lambda$ distribution around the peak that we will find , which is the main point of the paper.

\section{Results and discussions}
\begin{figure}
  \centering
  \includegraphics[width=0.8\textwidth]{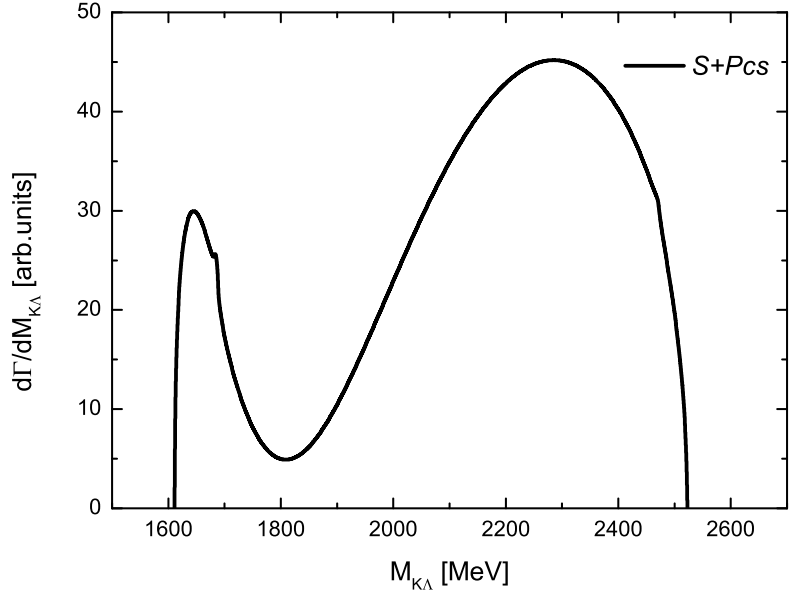}\\
  \caption{ $K^{0}\Lambda$ mass distribution with the amplitude of Eq.~(\ref{amplitude3}), where $s$ denotes the contribution of the
  contact and $s$-wave $K \Lambda$ interaction, Eq.~(\ref{amplitude1}), and $P_{cs}$ that of
  the hidden-charm state with strangeness,  Eq.~(\ref{amplitude2}).}\label{KLambda_S_Pc}
\end{figure}

We present here the results. In Fig.~\ref{KLambda_S_Pc} we show the $K^{0}\Lambda$ mass distribution taking into account the amplitude of Eq.~(\ref{amplitude3}).
Note that this amplitude only contains s-wave for both the $K^{0}\Lambda$ scattering and the $J/\psi \Lambda$ one. With the chiral unitary approach that we use, the $N^{*}(1535)(1/2^{-})$ resonance can
be dynamically generated, which is below the $K^{0}\Lambda$ threshold, but still will show up with an enhancement of the $K^0\Lambda$ mass distribution close to threshold. Yet, there will be contributions from other $N^{*}$ resonances which are not generated by the approach of Ref.~\cite{Inoue:2001ip}. We can take into account the effect of some relevant resonances, only to show that they do not affect the peak of the $J/\psi \Lambda$ mass distribution.

\begin{figure}
  \centering
  \includegraphics[width=0.8\textwidth]{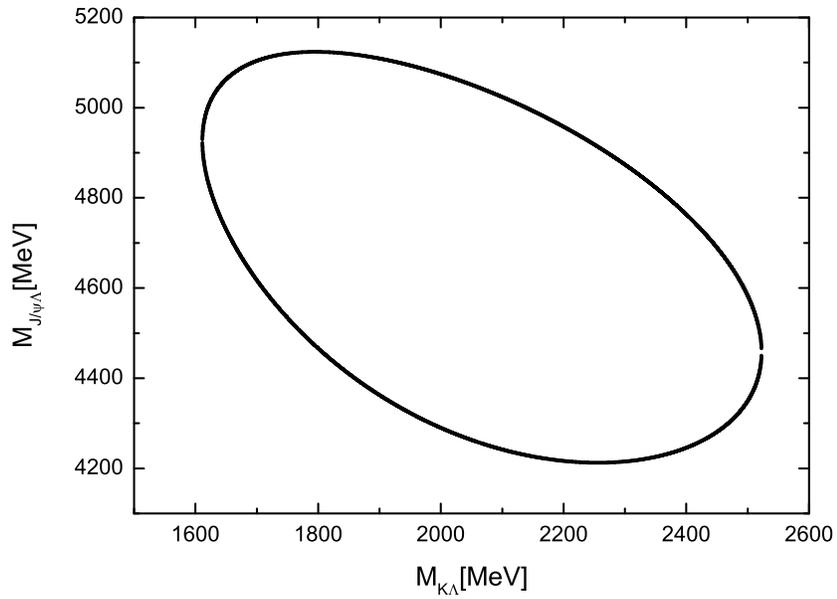}\\
  \caption{Dalitz plot for $\Lambda_{b}\rightarrow J/\psi K^0 \Lambda$.}\label{dalitz}
\end{figure}
By looking at the Dalitz plot of the process, shown in Fig.~\ref{dalitz}, we see that resonances decaying into  $K^{0}\Lambda$ with a mass less than 1750 MeV only contribute to $J/\psi \Lambda$ invariant masses beyond the peak in the $J/\psi \Lambda$ mass around 4650 MeV. Hence we only consider $N^{*}$ resonances from the PDG~\cite{Agashe:2014kda} above 1750 MeV with some coupling to $K^{0}\Lambda$. None of them has a sizeable coupling, and the few branching ratios are of the order of 10\% or less with large uncertainties. In view of this we take an extreme position of considering a couple of resonances, associating them a role in the $K^{0}\Lambda$ mass distribution rather sizeable, simply to see that, even then, there are no visible effects in the behavior of the peak of the $J/\psi\Lambda$ mass distribution.

\begin{figure}
  \centering
  \includegraphics[width=0.8\textwidth]{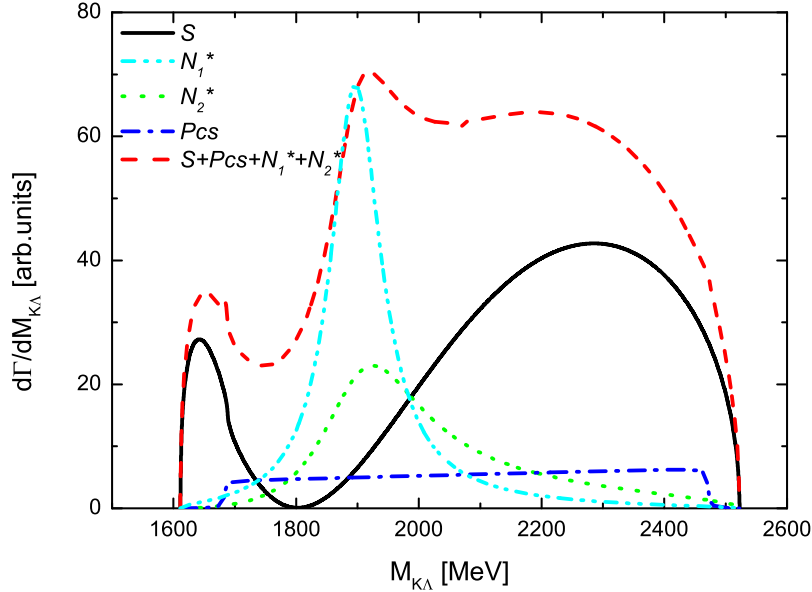}\\
  \caption{Effect of including the $N^{*}(1895)$ and $N^{*}(1900)$ resonances in $K^{0}\Lambda$ mass distribution, where $s$ denotes the contribution of the
  contact and $s$-wave $K \Lambda$ interaction, Eq.~(\ref{amplitude1}), $P_{cs}$ that of
  the hidden-charm state with strangeness,  Eq.~(\ref{amplitude2}), $N^{*}_1$ that of the $N^*(1895)$, and $N^{*}_2$ that of the $N^*(1900)$.}\label{FULL_KLambda2}
\end{figure}

Thus, we consider the contributions of  two resonances. The first one is the $N^{*}(1895)$ $(1/2^{-})$, and the other one the $N^{*}(1900)$ $(3/2^+)$. The first one, a two star resonance, according to Ref.~\cite{Anisovich:2011fc}, has a branching fraction to $K\Lambda$ of $18\pm 5\%$. The $N^{*}(1900)$ is a three star resonance and has a branching fraction to $K\Lambda$ of $0\thicksim10\%$ according to Ref.~\cite{Agashe:2014kda} and $16\pm 5\%$ according to Ref.~\cite{Anisovich:2011fc}. The $N^{*}(1895)(1/2^{-})$ has the same quantum numbers as the $N^{*}(1535)$ and its contribution will add coherently with no spin nor angular dependence. Hence we add to the amplitude $\mathcal{M}$ of  Eq.~(\ref{amplitude3}) the term
\begin{equation}\label{amplitudeN1895}
  T_{N^{*}(1895)}=\frac{\alpha M_{N^{*}(1895)}}{M_{K \Lambda}-M_{N^{*}(1895)}+{\rm i}\Gamma_{N^{*}(1895)}/2}
\end{equation}
and we tune the parameter $\alpha$ to get a sizeable effect in the $K^{0}\Lambda$ mass distribution. This is shown in Fig.~\ref{FULL_KLambda2}.

The effect of the $N^{*}(1900)$ is taken into account in a different way, because this one, having different quantum numbers, adds incoherently to $d\Gamma/d M_{K \Lambda}$. This is shown explicitly in the appendix. In this case we take into account the contribution of this resonance by substituting $|\mathcal{M}|^{2}$ of Eq.~(\ref{amplitude3}) by
\begin{equation}\label{amplitudeN1900}
    |\mathcal{M}|^{2} \rightarrow |\mathcal{M}|^{2}+ \beta \tilde{P}_{K}^{2}\left|\frac{1}{M_{K \Lambda}-M_{N^{*}(1900)}+{\rm i}\Gamma_{N^{*}(1900)}/2}\right|^{2},
\end{equation}
where $\tilde{P_{K}}$ is the $K^0$ momentum in the $K^0\Lambda$ rest frame to take into account that a $3/2^{+}$ state requires $K^0\Lambda$ in $p$-wave.
 The effect of introducing this resonance can be seen in the $K^{0}\Lambda$ mass distribution of Fig.~\ref{FULL_KLambda2}. As we have mentioned, the couplings $\alpha$, $\beta$ of these two resonances are taken such as to have a relatively large effect of these two resonances in the $K^{0}\Lambda$ mass distribution.

\begin{figure}
  \centering
  \includegraphics[width=0.6\textwidth]{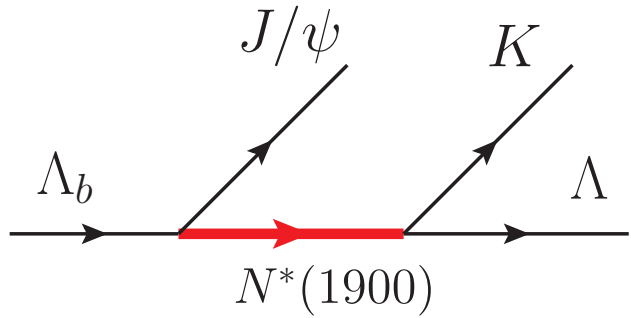}\\
  \caption{Diagrammatic representation of $\Lambda_b\rightarrow J/\psi K\Lambda$ via the intermediate production of a $3/2^+$ resonance.}\label{N1900}
\end{figure} 
The addition of the contribution of the $N^*(1900)$ ($3/2^+$) incoherently in Eq.~(\ref{amplitudeN1900}) requires a justification given in the appendix. The process of production of this resonance is depicted in Fig.~\ref{N1900}.
The vertex $\Lambda_b J/\psi R$ requires $1/2^++1^-\rightarrow 3/2^+$, which can proceed via $s$-wave, and parity can be changed by the weak interaction. The vertex $N^*(1900)\rightarrow K\Lambda$ is of the type $\vec{S} \cdot \vec{q}$, with $\vec{S}$ the spin transition operator from spin $3/2$ to spin $1/2$, and $\vec{q}$ the momentum of the kaon. Certainly, this has an angular dependence for a given polarization of the $\Lambda$ in the amplitude. 
To calculate $|\mathcal{M}|^2$, we must sum the amplitudes coherently with their spin structure, and the interference term brings extra angular dependence. However, we shall show explicitly in the appendix that the interference term between the amplitude of Eq.~(\ref{amplitude3}) and the one of $N^*(1900)$ vanishes when we sum $|\mathcal{M}|^2$ over the polarization of all the particles with spin. Even then, an angular dependence remains in the Dalitz plot because $M^2_{\rm inv}(J/\psi\Lambda)$ is proportional to ${\rm cos}\theta$, with $\theta$ the angle between $J/\psi$ and $K^0$ in the $K^0\Lambda$ rest frame (see appendix).

\begin{figure}
  \centering
  \includegraphics[width=0.8\textwidth]{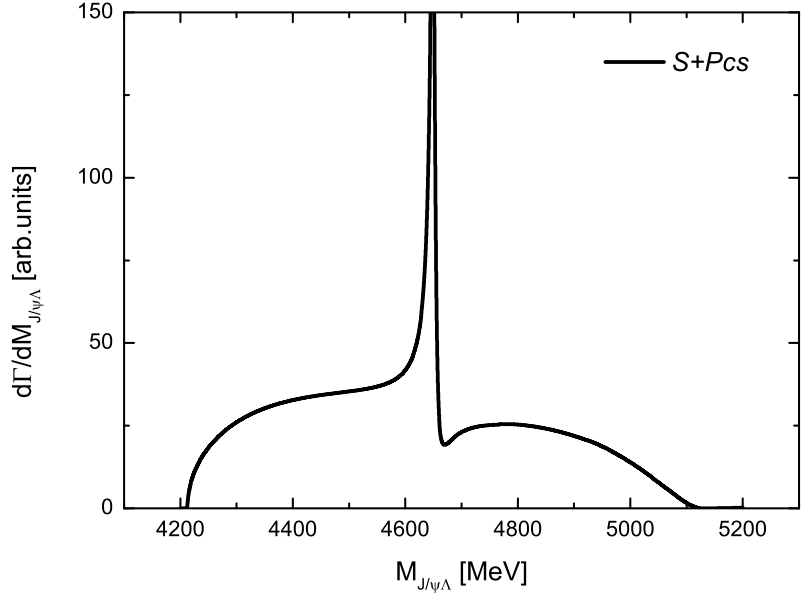}\\
  \caption{$J/\psi\Lambda$ mass distribution with the amplitude of Eq.~(\ref{amplitude3}), where $s$ denotes the contribution of the
  contact and $s$-wave $K \Lambda$ interaction, Eq.~(\ref{amplitude1}), and $P_{cs}$ that of
  the hidden-charm state with strangeness,  Eq.~(\ref{amplitude2}).}\label{Jpsilambda_S_Pc}
\end{figure}

Next we go to the $J/\psi\Lambda$ mass distribution. In Fig.~\ref{Jpsilambda_S_Pc}, we show  this distribution corresponding to the amplitude $\mathcal{M}$ of Eq.~(\ref{amplitude3}). One observes a clear peak around 4650 MeV corresponding to the pole in the $t_{J/\psi\Lambda\rightarrow J/\psi\Lambda}$ amplitude, associated to the hidden-charm strange resonance of Refs.~\cite{Wu:2010jy,Wu:2010vk}.

\begin{figure}
  \centering
  \includegraphics[width=0.8\textwidth]{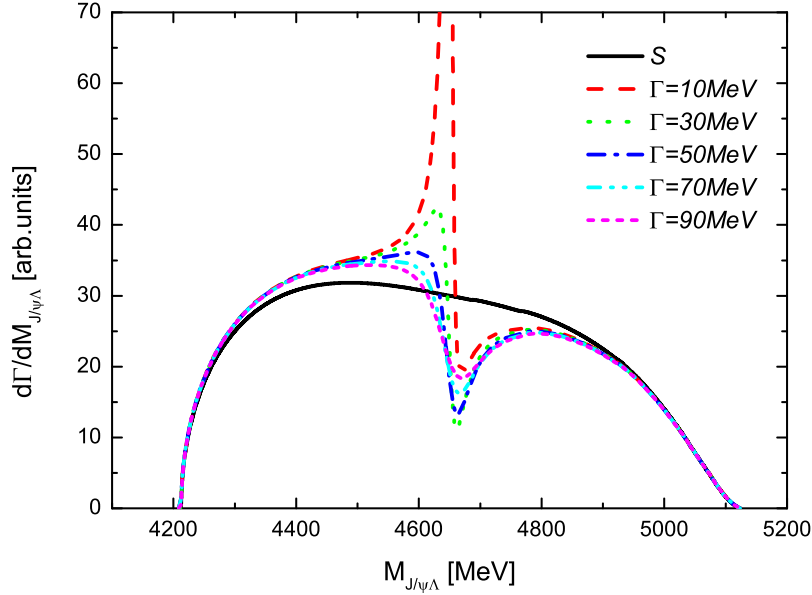}\\
  \caption{Effect of changing the width of the hidden-charm strange resonance in the $J/\psi\Lambda$ mass distribution, where $s$ denotes the contribution of the
  contact and $s$-wave $K \Lambda$ interaction, Eq.~(\ref{amplitude1}), and $\Gamma$ the width of
  the hidden-charm state with strangeness.}\label{Jpsilambda_S_Pc_Gamma}
\end{figure}
In Fig.~\ref{Jpsilambda_S_Pc_Gamma}, we show the effect of changing the width of the hidden-charm state with strangeness. One can observe that
as the width increases from about 10 MeV, predicted in Refs.~\cite{Wu:2010jy,Wu:2010vk}, the peak becomes less pronounced.  If the width is larger than 30 MeV, it seems that the existence of the
resonance can only be inferred from the interference effect, namely, a slight enhancement followed by a dip.

\begin{figure}
  \centering
  \includegraphics[width=0.8\textwidth]{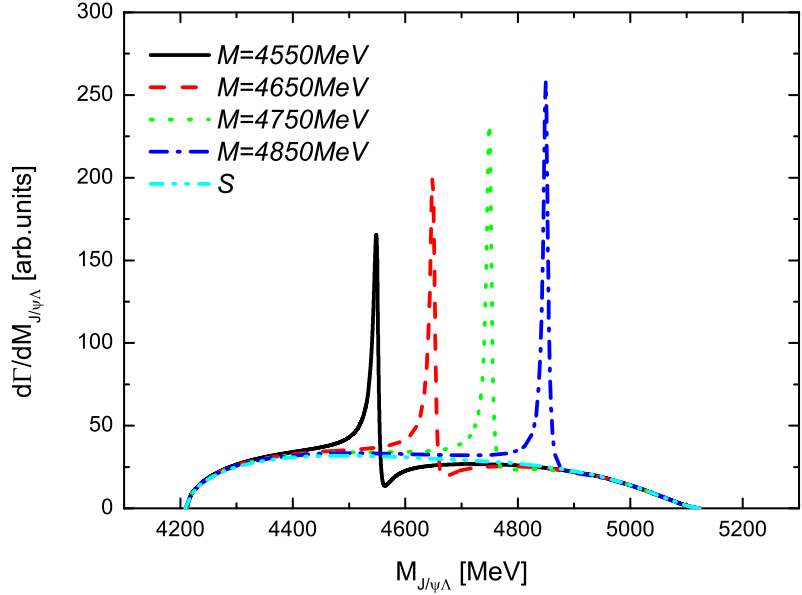}\\
  \caption{Effect of changing the mass of the hidden-charm strange resonance in the $J/\psi\Lambda$ mass distribution, where $s$ denotes the contribution of the
  contact and $s$-wave $K \Lambda$ interaction, Eq.~(\ref{amplitude1}), and $M$ the mass of
  the hidden-charm state with strangeness.}\label{Jpsilambda_S_Pc_MR}
\end{figure}

The effect of changing the mass of the hidden-charm strange resonance can be seen in Fig.~\ref{Jpsilambda_S_Pc_MR}.  It is clear that, by changing the mass of the resonance, the peak position varies, but the relevant point here is that the peak always appears even if this mass is varied in a wide range of values.

\begin{figure}
  \centering
  \includegraphics[width=0.8\textwidth]{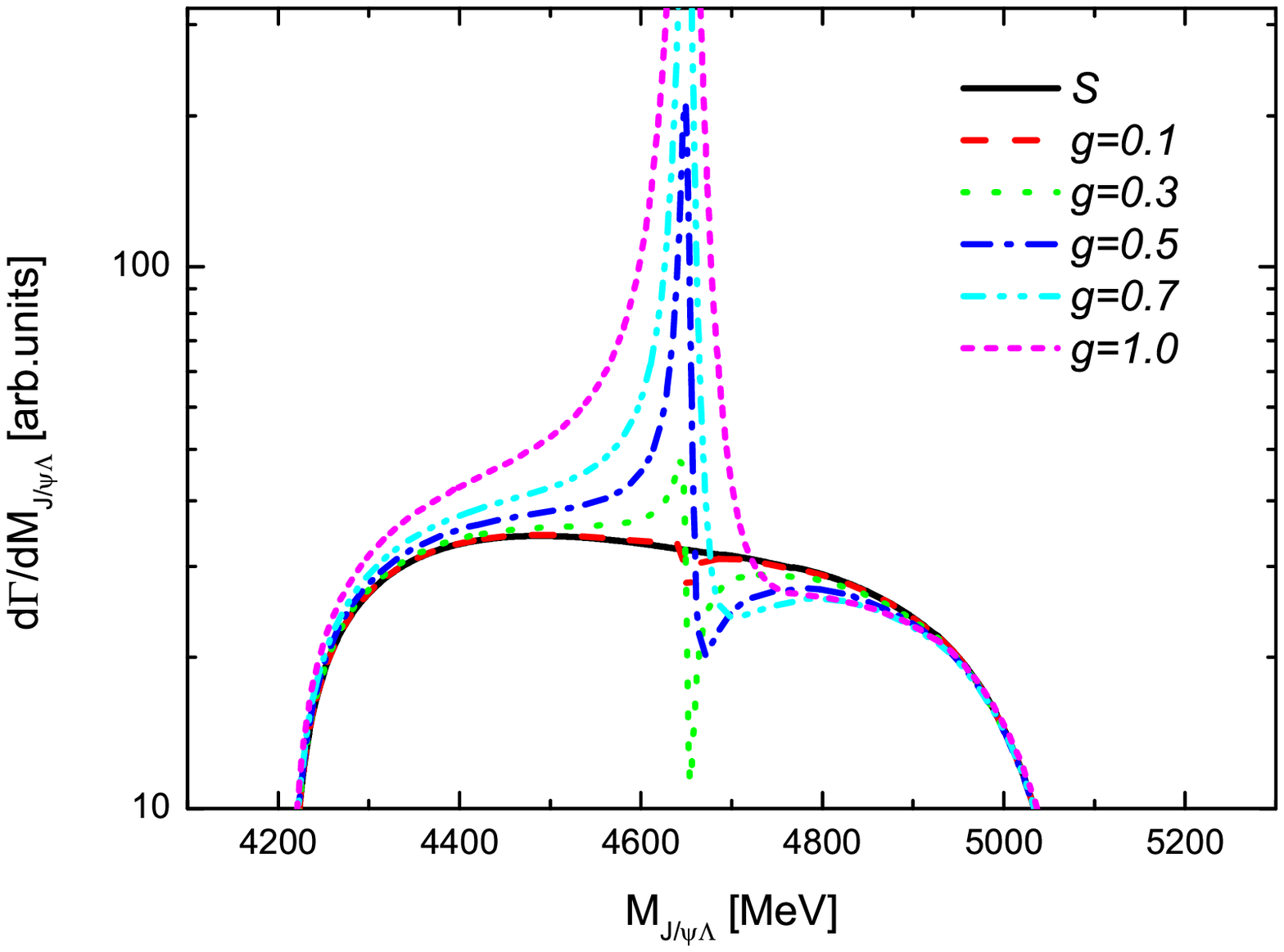}\\
  \caption{Effect of changing the coupling $g_{J/\psi\Lambda}$ of the hidden-charm strange resonance in the $J/\psi\Lambda$ mass distribution, where $s$ denotes the contribution of the
  contact and $s$-wave $K \Lambda$ interaction, Eq.~(\ref{amplitude1}), and $g$ the coupling to $J/\psi\Lambda$ of
  the hidden-charm state with strangeness.}\label{Jpsilambda_S_Pc_g}
\end{figure}

Another test that we perform is to see what happens when we change the coupling $g_{J/\psi\Lambda}$, as shown in Fig.~\ref{Jpsilambda_S_Pc_g}. We can see that the peak is rather stable and one can clearly see a structure even for values of $g_{J/\psi\Lambda}$ as low as $0.3$. 

\begin{figure}
  \centering
  \includegraphics[width=0.8\textwidth]{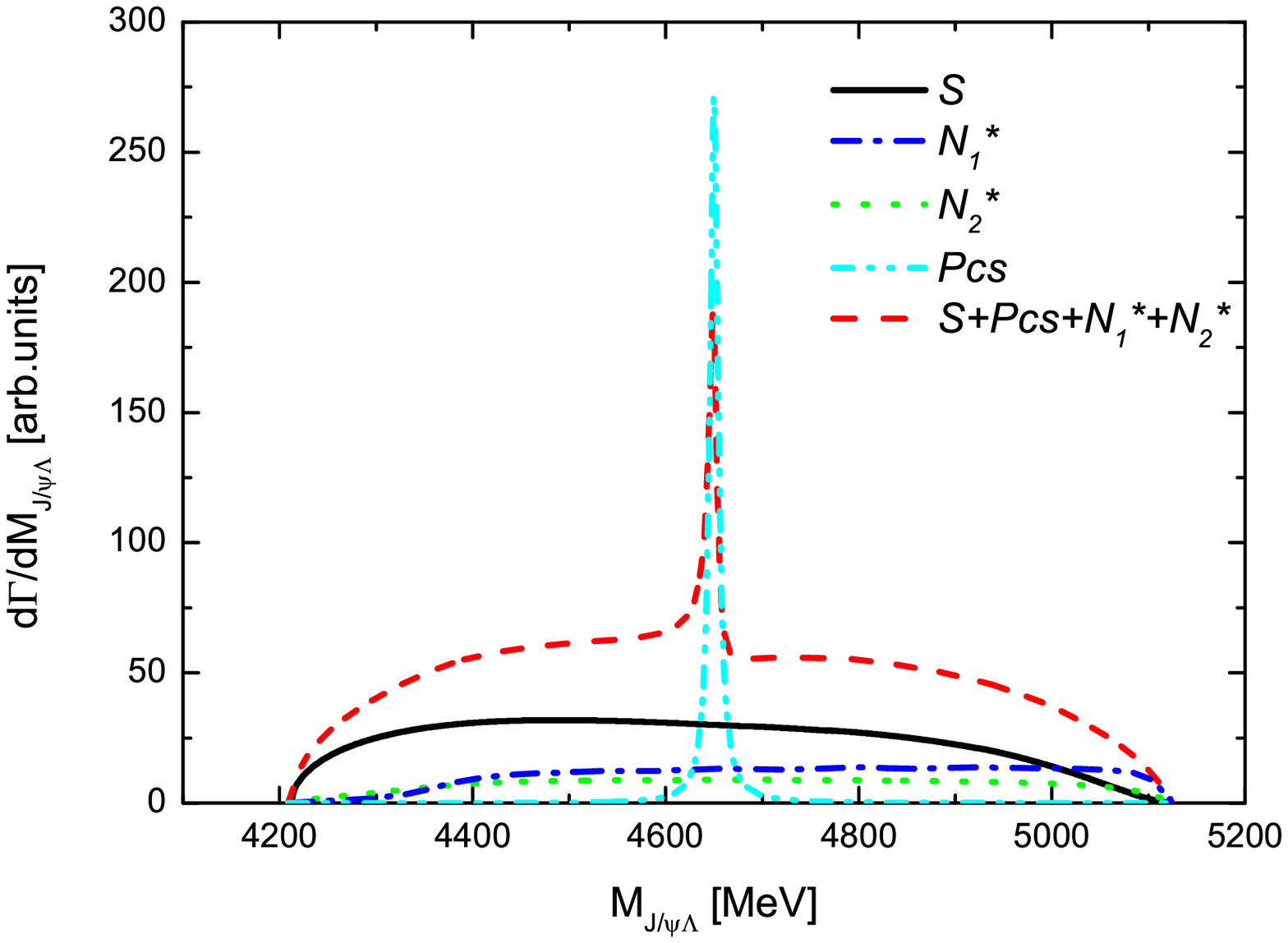}\\
  \caption{Effect of including the $N^{*}(1895)$ and $N^{*}(1900)$ resonances in $J/\psi\Lambda$ mass distribution, where $s$ denotes the contribution of the
  contact and $s$-wave $K \Lambda$ interaction, Eq.~(\ref{amplitude1}), $P_{cs}$ that of
  the hidden-charm state with strangeness,  Eq.~(\ref{amplitude2}), $N^{*}_1$ that of the $N^*(1895)$, and $N^{*}_2$ that of the $N^*(1900)$.}\label{FULL_JpsiLambda2}
\end{figure}

Next we want to see the effect of the extra resonances $N^{*}(1895)$ and $N^{*}(1900)$ in the $J/\psi\Lambda$ mass distribution. This is shown in Fig.~\ref{FULL_JpsiLambda2}.  It is clear that the consideration of the $N^{*}(1895)$ and $N^{*}(1900)$ resonances leads to a larger $J/\psi\Lambda$ mass distribution to match the increase found in the $K^{0}\Lambda$ mass distribution, but what is important for us is that the structure of the peak is not spoiled by the consideration of these or other possible resonances. The conclusion is then that, should there be a resonance with strangeness  and a reasonable coupling to $J/\psi\Lambda$ in s-wave in the region of masses studied here, its observation in the $J/\psi\Lambda$ mass spectrum of the $\Lambda_{b}\rightarrow J/\psi K^{0}\Lambda$ decay would be inevitable.

\section{The case of $J/\psi\Lambda$ in $3/2^-$}\label{sec:32}
So far we have only discussed the case where the system $J/\psi\Lambda$ is in $1/2^-$, which are possible quantum numbers of the state favored in Refs.~\cite{Wu:2010jy,Wu:2010vk} that we suggest to look for. We can discuss what would happen if one had $3/2^-$ which is also possible in Refs.~\cite{Wu:2010jy,Wu:2010vk}. In this case, as discussed earlier, we need a $p$-wave to match the $1/2^+$ of the $\Lambda_b$. 

For the amplitude of the tree level contribution of $p$-wave involving the spin explicitly like in the $s$-wave, we have,
\begin{equation}
T_{\rm tree}^{p-\rm wave}={\rm i}B\,\epsilon_{ijk}\,\sigma_k\, q_i \,\epsilon_j, \label{eq:ptree1}
\end{equation} 
and $q_i$ can be either the $J/\psi$ or $K^0$ momentum. However, if we take the loop of Fig.~\ref{Fig4} with one vertex in $p$-wave in the $J/\psi$ momentum and the $J/\psi \Lambda$ $t$-matrix in $s$-wave, as we have in Refs.~\cite{Wu:2010jy,Wu:2010vk}, the loop function vanishes. If the peak seen in Refs.~\cite{Aaij:2015tga,Aaij:2014zoa} corresponds to the molecule of Refs.~\cite{Wu:2010jy,Wu:2010vk} in $3/2^-$, then one needs the structure of Eq.~(\ref{eq:ptree1}) with  $q_i$ being the kaon momentum which we will take in the rest frame of the $K^0\Lambda$ system. 

For the amplitude of $J=3/2$ $J/\psi\Lambda$ final state interaction, we only have two structures, $\vec{q}\cdot \vec{\epsilon}$ and ${\rm i}(\vec{\sigma}\times \vec{q}\,)\cdot \vec{\epsilon}$, and we take  a combination of the two structures which is orthogonal to $\vec{\sigma}\cdot \vec{\epsilon}$, which projects $J/\psi\Lambda$ in $J=1/2$. We take the transition operator for $J=3/2$ $J/\psi\Lambda$ production as,
\begin{equation}
S_{3/2}= \left\langle m_{\Lambda} \mid  q_j\, \epsilon_j + {\rm i}b \,\epsilon_{ijk} \,\sigma_k \, q_i \epsilon_j \mid m_{\Lambda_b} \right\rangle, 
\end{equation}  
where the parameter $b$ can be obtained by the orthogonal relation,
\begin{equation}
\sum_{\chi,m_\Lambda}  \left\langle m_{\Lambda_b} |\vec{\sigma}\cdot \vec{\epsilon}\, |\chi_{J/\psi} m_\Lambda \right\rangle \left\langle \chi_{J/\psi}m_\Lambda \mid  q_i\, \epsilon_j + {\rm i}b \,\epsilon_{ijk} \,\sigma_k \, q_i \epsilon_j \mid m_{\Lambda_b} \right\rangle \equiv 0, \label{eq:b_orth}
\end{equation}
and we sum over the polarizations of $\Lambda$ ($m_\Lambda$) and $J/\psi$ ($\chi$). Recall $\sum_\chi \epsilon_i \epsilon_j =\delta_{ij}$ [see Eq.~(\ref{eq:2eps}) in the appendix]. Eq.~(\ref{eq:b_orth}) then leads to $b=1/2$. Thus, the transition operator for $J=3/2$ $J/\psi\Lambda$ production can be rewritten as,
\begin{equation}
S_{3/2}\equiv \left\langle m_{\Lambda} \mid  q_j\, \epsilon_j + \frac{\rm i}{2} \,\epsilon_{ijk} \,\sigma_k \, q_i \epsilon_j \mid m_{\Lambda_b} \right\rangle. 
\end{equation}  
And the contribution of the $J=3/2$ $J/\psi\Lambda$ final state interaction is,
\begin{equation}
T_{J/\psi\Lambda-3/2 }^{p-\rm wave} \varpropto B G_{J/\psi\Lambda}\,t_{J/\psi\Lambda,J/\psi\Lambda} \times S_{3/2}. \label{eq:pwaveJpsi}
\end{equation}

It is convenient to separate any operator into $p$-wave $J/\psi \Lambda$ spin $3/2$ and $1/2$. And for the operator for the $p$-wave $J/\psi \Lambda$ spin $J=1/2$, we also take a combination of the above two structures, $\vec{q}\cdot \vec{\epsilon}$ and ${\rm i}(\vec{\sigma}\times \vec{q}\,)\cdot \vec{\epsilon}$,
\begin{equation}
S_{1/2}= \left\langle m_{\Lambda} \mid  q_j\, \epsilon_j + {\rm i}b' \,\epsilon_{ijk} \,\sigma_k \, q_i \epsilon_j \mid m_{\Lambda_b} \right\rangle. 
\end{equation}  
Imposing $S_{1/2}$ to be orthogonal to $S_{3/2}$, we have,
\begin{equation}
\sum_{\chi,m_\Lambda}  \left\langle m_{\Lambda_b} |q_j\, \epsilon_j -{\rm i}b' \,\epsilon_{ijk} \,\sigma_k \, q_i \epsilon_j \, |\chi_{J/\psi} m_\Lambda \right\rangle \left\langle \chi_{J/\psi} m_\Lambda \mid  q_j\, \epsilon_j + \frac{\rm i}{2} \,\epsilon_{ijk} \,\sigma_k \, q_i \epsilon_j \mid m_{\Lambda_b} \right\rangle \equiv 0.
\end{equation}
According to this orthogonal relation, we obtain $b'=-1$. So, the operator for $J=1/2$ $J/\psi\Lambda$ can be expressed as,
\begin{equation}
S_{1/2}= \left\langle m_{\Lambda} \mid  q_j\, \epsilon_j - {\rm i} \,\epsilon_{ijk} \,\sigma_k \, q_i \epsilon_j \mid m_{\Lambda_b} \right\rangle. 
\end{equation}  

It is easy to separate the operator $T_{\rm tree}^{p-\rm wave}$ of Eq.~(\ref{eq:ptree1}) into $p$-wave $J/\psi\Lambda$ spin $3/2$ and $1/2$, which is,
\begin{equation}
T_{\rm tree}^{p-\rm wave} =\frac{2}{3}B\, S_{3/2} - \frac{2}{3}B\, S_{1/2}. \label{eq:pwavetree}
\end{equation}  
In addition, we have another structure for the $K^0\Lambda$ system in $3/2^+$ $N^*(1900)$ excitation, which is given by Eq.~(\ref{eq:t1900}),
\begin{equation}
T_{\rm N(1900)}^{p-\rm wave}=C \left\langle m_{\Lambda} \mid  q_i \left(\delta_{ij}-\frac{\rm i}{3} \epsilon_{ijk} \sigma_k  \right) \epsilon_j \mid m_{\Lambda_b} \right\rangle.
\end{equation}
We would like to separate it into $S_{3/2}$ and $S_{1/2}$, which can be easily expressed as,
\begin{equation}
T_{\rm N(1900)}^{p-\rm wave}= \frac{4}{9}\,C\, S_{3/2} + \frac{5}{9}\,C \, S_{1/2}. \label{eq:pwaveN1900}
\end{equation}

Finally, the full amplitude for the $\Lambda_{b}\rightarrow J/\psi K^{0}\Lambda$ process can be written as
\begin{eqnarray}
\mathcal{M}'&=& \mathcal{M}^{s-\rm wave} + \mathcal{M}^{p-\rm wave} \nonumber \\
&=& \left( T^{s-\rm wave}_{\rm tree} + T^{s-\rm wave}_{K^0\Lambda} + T^{s-\rm wave}_{N1895} \right) + 
T_{\rm tree}^{p-\rm wave} + T_{J/\psi\Lambda-3/2 }^{p-\rm wave} + T_{\rm N(1900)}^{p-\rm wave}, \label{eq:amp_pwave}
\end{eqnarray} 
where the terms $T^{s-\rm wave}_{\rm tree}$ and  $T^{s-\rm wave}_{K^0\Lambda}$ are the same as the terms $T_{\rm tree}$ and $T_{K^0\Lambda}$
involving the $s$-wave coupling $\left\langle m_\Lambda |\vec{\sigma}\cdot \vec{\epsilon}\, |M_{\Lambda_b} \right\rangle$. With  Eqs.~(\ref{eq:pwavetree}), (\ref{eq:pwaveJpsi}) and (\ref{eq:pwaveN1900}), the full amplitude can be rewritten as,
\begin{eqnarray}
\mathcal{M}'&=& V'_p\left[ h_{K^0\Lambda} + \sum_{i}h_{i}G_{i}t_{i, K^{0} \Lambda}+\frac{\alpha' N^{*}(1895)}{M_{K \Lambda}-M_{N^{*}(1895)}+i\Gamma_{N^{*}(1895)}/2} \right] \left\langle m_\Lambda |\vec{\sigma}\cdot \vec{\epsilon}\, |m_{\Lambda_b} \right\rangle \nonumber \\
&& + B \left[ \left( \frac{2}{3} + \frac{2}{3} G_{J/\psi\Lambda}t_{J/\psi\Lambda,J/\psi\Lambda} + \frac{4}{9} \frac{\beta' N^{*}(1900)}{M_{K \Lambda}-M_{N^{*}(1900)}+{\rm i}\Gamma_{N^{*}(1900)}/2} \right) \times S_{3/2} \right. \nonumber \\ && \left. + \left( -\frac{2}{3} +\frac{5}{9} \frac{\beta' N^{*}(1900)}{M_{K \Lambda}-M_{N^{*}(1900)}+i\Gamma_{N^{*}(1900)}/2}\right) \times S_{1/2}   \right], \label{eq:amp_SP}
\end{eqnarray}
where $\alpha'$ and  $\beta'=C/B$ are the overall normalization factors of the contribution of $N^*(1895)$ and $N^*(1900)$, and we will take the value for them to make their contributions sizeable. The value of $B$ will be discussed later.

By looking at the appendix and summing and averaging over the polarization, we obtain,
\begin{eqnarray}
\bar{\sum}\sum\left| \mathcal{M}'\right|^2 &=& 3 V^{\prime 2}_p \left|  h_{K^0\Lambda} + \sum_{i}h_{i}G_{i}t_{i,K^{0} \Lambda} + \frac{\alpha' N^{*}(1895)}{M_{K \Lambda}-M_{N^{*}(1895)}+i\Gamma_{N^{*}(1895)}/2} \right|^2 \nonumber \\
&& + \frac{3}{2} B^2 \vec{q}^{\,2}\left|  \frac{2}{3} + \frac{2}{3} G_{J/\psi\Lambda} t_{J/\psi\Lambda,J/\psi\Lambda} + \frac{4}{9} \frac{\beta' N^{*}(1900)}{M_{K \Lambda}-M_{N^{*}(1900)}+{\rm i}\Gamma_{N^{*}(1900)}/2} \right|^2 \nonumber \\
&& + 3 B^2 \vec{q}^{\,2}\left| -\frac{2}{3} +\frac{5}{9} \frac{\beta' N^{*}(1900)}{M_{K \Lambda}-M_{N^{*}(1900)}+i\Gamma_{N^{*}(1900)}/2} \right|^2 . \label{eq:amp2_SP}
\end{eqnarray}

We have now a new parameter $B$, which is unrelated to $V'_p$, unlike in the case of $J=1/2$ where we had only one parameter $V_p$. 
We also see that the strange hidden charm resonance, present in our case in $t_{J/\psi\Lambda,J/\psi\Lambda}$, appears in the second term of Eq.~(\ref{eq:amp2_SP}) and interferes with the $p$-wave tree level and the $N^*(1900)$. Then, unlike the case of $s$-wave, where we can make a mapping of the interaction in other sectors, here one cannot because the explicit $N^*$ resonances of relevance to this case do not have a counterpart  in the $\Lambda^*$ resonances of relevance in the $\Lambda_b \to J/\psi K^- p$ reaction. To investigate what can happen in this case, we take the position to assume that both the $p$-wave tree level and the $N^*(1900)$ terms have a size of the same order of magnitude as the contribution of $s$-wave $K\Lambda$.

\begin{figure}
  \centering
  \includegraphics[width=0.8\textwidth]{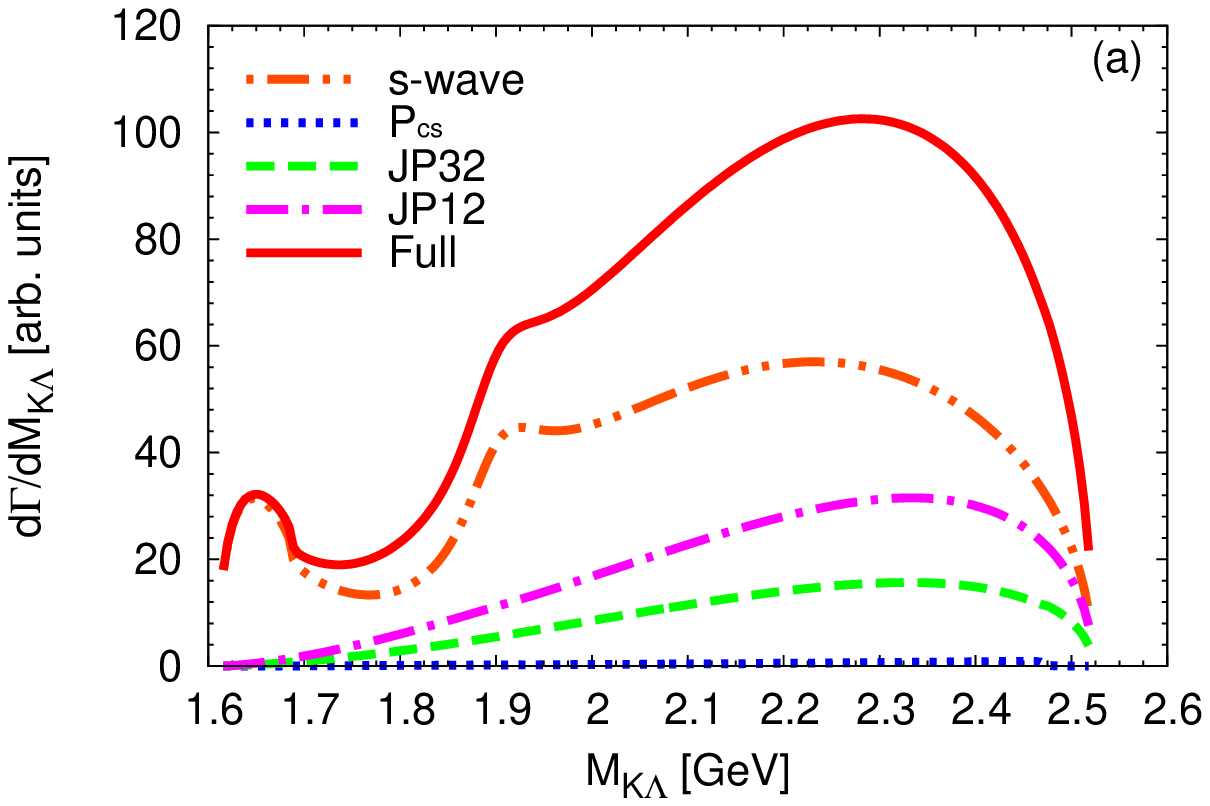}\\
  \includegraphics[width=0.8\textwidth]{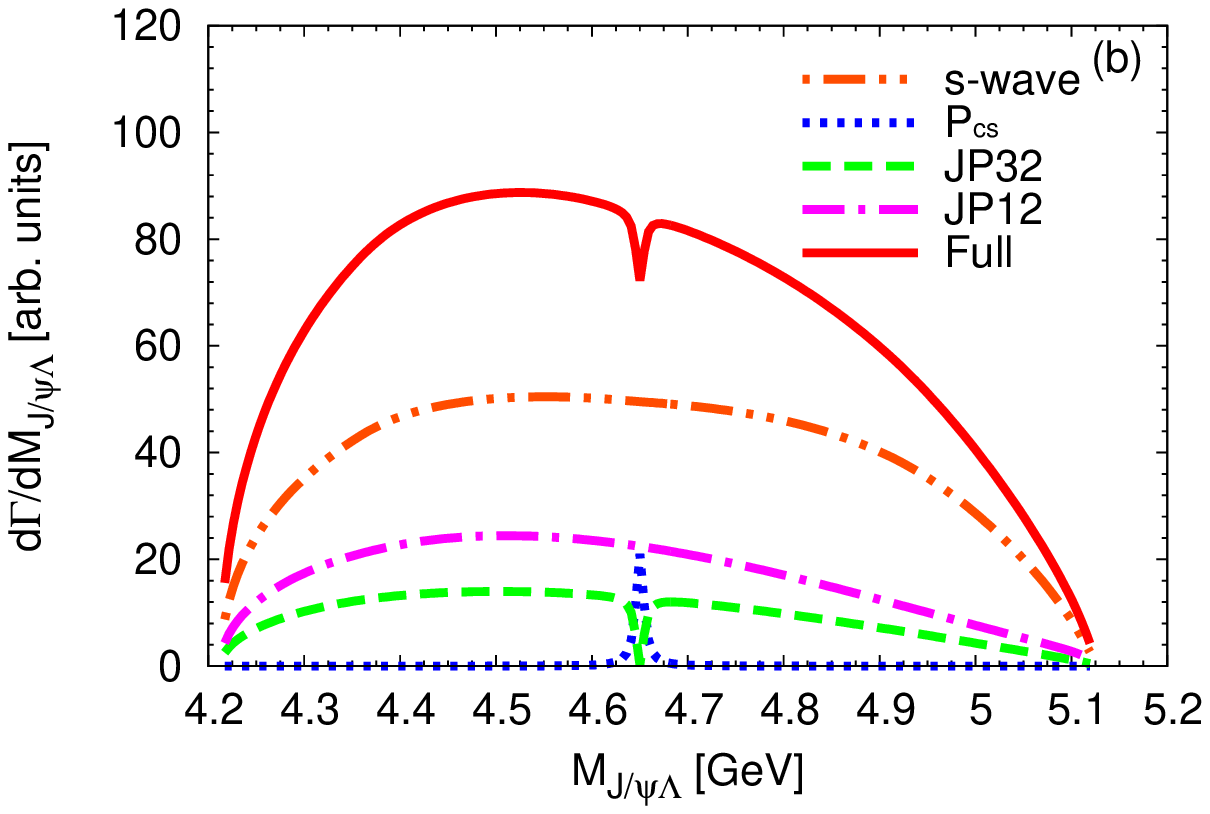}\\
  \caption{ $K^{0}\Lambda$ (a) and $J/\psi\Lambda$ (b) mass distributions with the amplitude of Eq.~(\ref{eq:amp_SP}). The curves labeled as `s-wave', `JP32' and `JP12' represent the contributions of the first, second and third terms of Eqs.~(\ref{eq:amp_SP}) and (\ref{eq:amp2_SP}), and the `$P_{cs}$' curves shows the contribution of the hidden-charm state with strangeness, finally the `Full' curves give the results for the all the contribution as shown in Eqs.~(\ref{eq:amp_SP}) and (\ref{eq:amp2_SP}).}\label{KLambda_S_Pc_threehalf}
\end{figure}

With this input, we perform the calculations, and evaluate the $K^0\Lambda$ and $J\psi \Lambda$ invariant mass distributions in Fig.~\ref{KLambda_S_Pc_threehalf}. As said above, we have again played with changing the strength of the $N^*(1895)$ and $N^*(1900)$ ($\alpha'$ and $\beta'$), giving them a sizeable contribution, in spite of which the signal is clearly seen.  As we can see, instead of a peak, there is now a dip structure, coming from the interference of the $T_{\rm tree}^{p-\rm wave}$ and $T_{J/\psi\Lambda-3/2 }^{p-\rm wave}$. Note that this behavior is relatively common in hadron physics. For example, the $f_0(980)$ manifests itself as a clear peak in the $\pi^+\pi^-$ invariant mass of the $J/\psi  \to \phi \pi^+\pi^-$~\cite{Wu:2001vz,Augustin:1988ja} and $B_s \to J/\psi \pi^+\pi^-$~\cite{Aaij:2011fx} reactions, but shows up as a dip in the $s$-wave $\pi\pi$ scattering amplitude~\cite{Pelaez:2015qba}. In addition, we also found out that the pentaquark $P_c(4450)$  gives a dip structure in the the $J/\psi p$ invariant mass of the $\Lambda_b\to J/\psi p \pi^-$\cite{Wang:2015pcn}, even assuming an $s$-wave production mechanism.

\section{Conclusion}
We have performed a theoretical study of the $\Lambda_{b}\rightarrow J/\psi K^{0}\Lambda$ reaction with the aim of making predictions for the possible observation of a baryon state of hidden-charm with strangeness that was predicted in Refs.~\cite{Wu:2010jy,Wu:2010vk}. We take into account the most important mechanisms in the $K^{0}\Lambda$ interaction that can interfere with the production amplitude for this molecular state, and we find that with the couplings of the resonance to the $J/\psi\Lambda$ channel found in Refs.~\cite{Wu:2010jy,Wu:2010vk}, and even one half of that, one obtains a clear peak in the $J/\psi\Lambda$ mass distribution. We played with uncertainties, changing the mass of the resonance and its coupling to $J/\psi\Lambda$, and we found the peak signal rather stable.  The peak remained distinguishable even after we introduced the effect of other resonances which do not appear in our theoretical framework. 
The study was done explicitly for the case where the $J/\psi\Lambda$ system would be in $s$-wave with $J^P=1/2^-$, and also for the case that the state would be a $3/2^-$, which was driven by a primary $\Lambda_b\to J/\psi K^0\Lambda$ vertex involving $p$-wave for the kaon. 
In this latter case, the predictions are more uncertain, but we still see a signal of the strange hidden charm state, through a negative interference with other terms. 
 In view of this, and the fact that the $\Lambda_{b}\rightarrow J/\psi \pi^{-}p$ decay has already been observed and is under reanalysis by the LHCb group at present~\cite{Sheldon}, we can only encourage to look also for the $K^{0}\Lambda$ channel, which is a coupled channel to the $\pi^{-}p$, in order to eventually observe this promising ``sister" of some of the resonances observed in Ref.~\cite{Aaij:2015tga}.

\section{Acknowledgements}

One of us, E. Oset, wishes to acknowledge support from the Chinese Academy of Science in the Program of Visiting Professorship for Senior International Scientists (Grant No. 2013T2J0012).
This work is partly supported by the National Natural Science Foundation of China under Grant Nos. 1375024, 11522539, 11505158, 11475015, and 11475227, the Spanish Ministerio de Economia y Competitividad and European FEDER funds under the contract number FIS2011-28853-C02-01 and FIS2011-28853-C02-02, and the Generalitat Valenciana in the program Prometeo II-2014/068.  This work is also supported by  the China Postdoctoral Science Foundation (No. 2015M582197), the Postdoctoral Research Sponsorship in Henan Province (No. 2015023) and the Open Project Program of State Key Laboratory of Theoretical Physics, Institute of Theoretical Physics, Chinese Academy of Sciences, China (No. Y5KF151CJ1).   
 
\appendix 
\section{Evaluation of the $\bar \sum\sum |t|^2$ for the contribution of the $N^*(1900)$}
\label{appendix}
In this appendix we give details on the evaluation of the matrix elements of the mechanisms involved and their interference in the double differential mass distribution of Eq.~(\ref{distribution}). As is well known, the integrated width over the phase space is given by,
\begin{equation}
\Gamma=\int\frac{{\rm d}^3p_J}{(2\pi)^3}\frac{1}{2\omega_J}\frac{1}{8\pi^2}\int{\rm d}\hat{\Omega}(\hat k)M_\Lambda \tilde{k}\frac{1}{M_{\rm inv}}\bar{\sum}\sum|t|^2, \label{eq:width}
\end{equation}
where $M_{\rm inv}=\sqrt{(P_K+P_\Lambda)^2}$,  and $\tilde{k}$ is the momentum of the kaon in the rest frame of $K^0\Lambda$ system. The angular integration over the solid angle ${\rm d} \hat{\Omega}(\hat{k})$ is also done in the $K^0\Lambda$ rest frame, and $\vec{p}_J$, $\omega_J$ refer to the momentum and energy of the $J/\psi$. The $\vec{p}_J$ integration is usually done in the $\Lambda_b$ rest frame.

We now assume that the $\Lambda_b\to J/\psi K^0\Lambda$ proceeds with the lowest possible angular momentum, in this case $L=0$, yet we must construct a scalar involving the polarization, $\vec{\epsilon}$, of the $J/\psi$. With $K^0$ and $\Lambda$ also in $s$-wave, as we have in our main mechanism, the suited operator is $\vec{\sigma}\cdot \vec{\epsilon}$. This operator projects the $J/\psi\Lambda$ system in $J=1/2$, as shown explicitly in Appendix B of Ref.~\cite{Garzon:2012np}, which is consistent with the discussion done earlier in Section~\ref{sec:formalism}. This means that $V_p$ in Eq.~(\ref{amplitude1}) implicitly contains this operator. Since both the $K^0\Lambda$ and $J/\psi\Lambda$ interactions are in $L'=0$ in our formalism, the structure of the terms involved in Eq.~(\ref{amplitude3}) is of the type~\footnote{Note that the $J/\psi\Lambda \to J/\psi\Lambda$ in $s$-wave in the local hidden gauge approach is of the type $\vec{\epsilon}\cdot \vec{\epsilon}^{\,\prime}$. Upon summing over the polarization of the $J/\psi$ in the loop of Fig.~\ref{Fig4}, $\sum_{\rm pol}\epsilon_j\epsilon_m\approx \delta_{jm}$ [see Eq.~(\ref{eq:2eps})] and the $\vec{\sigma}\cdot \vec{\epsilon}$ structure of the tree level remains for the rescattering term of Fig.~\ref{Fig4}.},
\begin{equation}
\mathcal{M}=A\left\langle m_\Lambda |\vec{\sigma}\cdot \vec{\epsilon}\, |M_{\Lambda_b} \right\rangle.
\label{eq:mswave}
\end{equation}
On the other hand, when we produce the $3/2^+$ state of the $N^*(1900)$, instead of the $\vec{\sigma}\cdot \vec{\epsilon}$ operator, we need now, also in $s$-wave, the coupling $\vec{S}^\dagger \cdot \vec{\epsilon}$, where $\vec{S}$ is the transition spin operator from spin $3/2$ to spin $1/2$. The coupling of this resonance to the final $K^0\Lambda$ is of the type $\vec{S}\cdot \vec{q}$, with $\vec{q}$ the $K^0$ momentum in the $K^0\Lambda$ rest frame. The amplitude corresponding  to the diagram of Fig.~\ref{N1900} would be given by,  
\begin{eqnarray}
t&=&B\sum_{M_R}\left\langle m_{\Lambda} |\vec{S}\cdot \vec{q} \,| M_R\right\rangle \left\langle M_R |\vec{S}^\dagger\cdot \vec{\epsilon} \, | m_{\Lambda_b} \right\rangle,
\end{eqnarray}
where $\vec{\epsilon}$ is the polarization of the $J/\psi$ and $M_R$ the one of the resonance. Using the sum over the polarization  of the $N^*(1900)$ resonance we have,
\begin{equation}
t=B\left\langle m_{\Lambda} \mid  q_i \left(\delta_{ij}-\frac{\rm i}{3} \epsilon_{ijk} \sigma_k  \right) \epsilon_j \mid m_{\Lambda_b} \right\rangle.\label{eq:t1900}
\end{equation}

All the $p$-wave structure were written in terms of $S_{1/2}$, $S_{3/2}$, which are orthogonal. By construction $S_{3/2}$ is also orthogonal to $\vec{\sigma}\cdot \vec{\epsilon}$. The operator $S_{1/2}$ is not orthogonal to  $\vec{\sigma}\cdot \vec{\epsilon}$ in the sense of Eq.~(\ref{eq:b_orth}), but when the sum and average over all polarizations is done [extra sum over $m_{\Lambda_b}$ in Eq.~(\ref{eq:b_orth})], the interference term also vanishes.
Hence we must just preform the sum and average over spins for $|\mathcal{M}|^2$, $|S_{3/2}|^2$ and $|S_{1/2}|^2$,
\begin{eqnarray}
\bar{\sum}\sum\vert \mathcal{M}\vert ^2&=& \frac{1}{2}|A|^2\sum_{m_\Lambda}\sum_{m_{\Lambda_b}}\sum_\chi \left\langle m_\Lambda  |\vec{\sigma}\cdot \vec{\epsilon}\,| m_{\Lambda_b}  \right\rangle \left\langle m_{\Lambda_b}  |\vec{\sigma}\cdot \vec{\epsilon}\,| m_{\Lambda}  \right\rangle \nonumber \\
&=& |A|^2 \sum_\chi \vec{\epsilon}\cdot\vec{\epsilon}=3|A|^2,
\end{eqnarray}
where $\chi$ stands for the polarization index of the $J/\psi$.
On the other hand,
\begin{eqnarray}
\bar{\sum}\sum |S_{3/2}|^2 &=& \frac{1}{2} \sum_{m_{\Lambda_b}}\sum_{m_\Lambda}\sum_{\chi} \left\langle m_{\Lambda} \left|q_i \epsilon_j  \left(\delta_{ij}+\frac{\rm i}{2} \epsilon_{ijk} \sigma_k  \right) \right| m_{\Lambda_b}\right\rangle \nonumber \\ && \times \left\langle m_{\Lambda_b} \left|q_l \epsilon_m \left(\delta_{lm}-\frac{\rm i}{2} \epsilon_{lmn} \sigma_n  \right)  \right| m_{\Lambda} \right\rangle \nonumber \\
&=& \frac{1}{2}\sum_{m_\Lambda}  \left\langle m_{\Lambda} \left| q_i q_l \left[ \delta_{il} +\frac{1}{4}\left(\delta_{li}\delta_{nk}-\delta_{lk}\delta_{ni}\right)\delta_{nk}\right] \right|m_{\Lambda} \right\rangle \nonumber \\
&=& \frac{1}{2}\cdot 2 \left[ \vec{q}^{\,2}+\frac{1}{4}\left(3\vec{q}^{\,2}-\vec{q}^{\,2}\right) \right] = \frac{3}{2}\vec{q}^{\,2},
\end{eqnarray} 
and,
\begin{eqnarray}
\bar{\sum}\sum |S_{1/2}|^2 &=& \frac{1}{2} \sum_{m_{\Lambda_b}}\sum_{m_\Lambda}\sum_{\chi} \left\langle m_{\Lambda} \left|q_i \epsilon_j  \left(\delta_{ij}-{\rm i} \epsilon_{ijk} \sigma_k  \right) \right| m_{\Lambda_b}\right\rangle \nonumber \\ && \times \left\langle m_{\Lambda_b} \left|q_l \epsilon_m \left(\delta_{lm}+{\rm i} \epsilon_{lmn} \sigma_n  \right)  \right| m_{\Lambda} \right\rangle \nonumber \\
&=& \frac{1}{2}\sum_{m_\Lambda}  \left\langle m_{\Lambda} \left| q_i q_l \left[ \delta_{il} +\left(\delta_{li}\delta_{nk}-\delta_{lk}\delta_{ni}\right)\delta_{nk}\right] \right|m_{\Lambda} \right\rangle \nonumber \\
&=& \frac{1}{2}\cdot 2 \left[ \vec{q}^{\,2}+\left(3\vec{q}^{\,2}-\vec{q}^{\,2}\right) \right] = 3\vec{q}^{\,2},
\end{eqnarray}
where we take,
\begin{equation}
\sum_\chi \epsilon_j \epsilon_m = \delta_{jm} +\frac{(p_{J/\psi})_j(p_{J/\psi})_m}{(M_{J/\psi})^2}\simeq \delta_{jm}, \label{eq:2eps}
\end{equation}
ignoring the small term $(p_{J/\psi})_j(p_{J/\psi})_m/{(M_{J/\psi})^2}$, but one can keep it and the conclusions are the same.

One should note that in the Dalitz distribution of Eq.~(\ref{distribution}), there is an implicit angular dependence. Indeed, the $J/\psi\Lambda$ invariant mass is related to the angle between $J/\psi$ and $K^0$   as shown below. We have,
\begin{eqnarray}
M^2_{\rm inv}(J/\psi\Lambda)&=& (P_{\Lambda_b}-P_K)^2 \nonumber \\
&=& M^2_{\Lambda_b} + m^2_K -2P_{\Lambda_b}P_K \nonumber \\ &=& M^2_{\Lambda_b} + m^2_K -2P^0_{\Lambda_b}P^0_K+2\vec{P}_{\Lambda_b}\cdot\vec{P}_K,
\end{eqnarray}
where $P_{\Lambda_b}$, $P_K$ are the four-momentum of the $\Lambda_b$ and the $K^0$. In the $K^0\Lambda$ rest frame, where we preform the angular integration ${\rm d}\hat{\Omega}$, one has $\vec{P}_{\Lambda_b}=\vec{P}_{J/\psi}$, since $\vec{P}_{\Lambda_b}-\vec{P}_{J/\psi}=\vec{P}_{K}+\vec{P}_\Lambda=0$. The values of $P^0_{\Lambda_b}$, $P_K^0$, $P_{J/\psi}$ and $P_K$ in that frame are all functions of $M_{\rm inv}(K^0\Lambda)$ and hence the term $\vec{P}_{\Lambda_b}\cdot\vec{P}_K$ becomes $\vec{P}_{J/\psi}\cdot \vec{P}_K$. Other angles in the integral of Eq.(\ref{eq:width}) can be easily done and one comes out with the standard formula of Eq.~(\ref{distribution}), which is provided in the PDG~~\cite{Agashe:2014kda} for the case when one sums over all the polarizations of the particles involved.

\end{document}